\documentclass[preprint,showpacs,preprintnumbers,amsmath,amssymb]{revtex4}

\usepackage{graphicx}   
\usepackage{dcolumn}    
\usepackage{bm}             

\newcommand{\bra}[1]{\langle #1|}
\newcommand{\ket}[1]{|#1\rangle}

\begin{document}

\title{Local-dephasing-induced entanglement sudden death in two-component finite-dimensional systems}
\author{Kevin Ann}
\email{kevinann@bu.edu}
\affiliation{Department of Physics, Boston University, Boston, MA 02215}

\author{Gregg Jaeger}
\email{jaeger@bu.edu}
\affiliation{Quantum Imaging Lab, Department
of Electrical and Computer Engineering, and Division of Natural
Sciences, Boston University, Boston, MA 02215\\}
\date{\today}

\begin{abstract}
Entanglement sudden death (ESD), the complete loss of entanglement
in finite time, is demonstrated to occur in a class of bipartite
states of qu-$d$-it pairs of any finite dimension $d>2$, when
prepared in so-called `isotropic states' and subject to multi-local
dephasing noise alone. This extends previous results for qubit pairs
[T. Yu, J. H. Eberly, Phys. Rev. Lett. {\bf 97}, 140403 (2006)] to
all qu-$d$-it pairs with $d>2$.
\end{abstract}

 \maketitle

Entanglement is perhaps the most quantum mechanical property a
physical system can possess. The behavior of entanglement under the
influence of environmental noise is important to quantum
measurements and enables powerful quantum computations
\cite{NC,Preskill}. Noise, even acting locally or on phases alone,
may cause not only state decoherence but also state disentanglement
\cite{YE02,YE03,YE04,AJ07,AJ072,DH04,YE06,CLR04,ADH07,LCD07}.
Indeed, recent work  has shown that even weak local noise acting on
bipartite states of infinite-dimensional systems, pairs of qubits,
and qubit--qutrit systems can lead to \emph{entanglement sudden
death} (ESD), a total loss of state entanglement in finite time with
generic decoherence taking place only asymptotically
\cite{DH04,YE06,ADH07,LCD07}. Here, we extend these results, showing
the existence of weak local dephasing noise induced ESD in bipartite
isotropic qudit-qudit states \cite{HH99} for every finite dimension
$d>2$ using the entanglement of formation $E_{\rm f}$ as a measure
of entanglement.

The isotropic states are those invariant under transformations of
the form $U \otimes U^{\ast}$, where $U$ is unitary \cite{HH99}. The
general $d\times d$-dimensional isotropic states $\rho_{\rm iso}(d)$
are convex combinations of a maximally mixed state
$(d^{-2})\mathbb{I}_{d^{2}}$ and a maximally entangled projector
$P(\ket{\Psi(d)})\equiv\ket{\Psi(d)}\bra{\Psi(d)}$:
\begin{equation}
\rho_{\rm iso}(d) = \left(\ \frac{1 - F}{d^{2} - 1} \right)
\mathbb{I}_{d^{2}} + \left(\ \frac{F d^{2} - 1}{d^{2} - 1} \right)
P(\ket{\Psi(d)}),
\end{equation}
where $d>1$, $\mathbb{I}_{d^{2}}$ is the $d^{2} \times d^{2}$
identity matrix, $\ket{\Psi(d)} = (1/\sqrt{d})\sum_{i =
1}^{d}{\ket{i} \ket{i}}$; the fidelity $F(\rho_{\rm iso}(d),
P(\ket{\Psi(d)})) ={\rm tr}\left(\rho_{\rm iso}(d)
P(\ket{\Psi(d)})\right)$ \cite{Jozsa}, which is bounded by 0 and 1
and appears self-consistently in the formal definition of isotropic
states \cite{HH99}, proves convenient for our study of
disentanglement. The state $\rho_{\rm iso}(d)$ is separable if and
only if $F\left(\rho_{\rm iso}(d),P(\ket{\Psi})\right)\leq F_{\rm
critical}(d)\equiv d^{-1}$, according to the standard measure of
entanglement, the entanglement of formation: for the isotropic
states $\rho_{\rm iso}(d)$ for $d>2$,
\begin{equation}
E_{\rm f}\left(\rho_{\rm iso}\right) =
\begin{cases}
0, &  F \leq \frac{1}{d}\ , \\
R_{1, d-1}\left( F \right), & F\in\left[\frac{1}{d}, \frac{4(d-1)}{d^{2}} \right]\ , \\
\frac{d \log(d-1)}{d-2}\left( F - 1 \right) + \log d, & F \in \left[
\frac{4(d-1)}{d^{2}}, 1\right]\ ,
\end{cases}
\end{equation}
where $R_{1, d-1}\left( F \right) = H_{2}\left(\xi(F)\right) +
\left[1 - \xi(F) \right]\log_{2}(d-1)$, \ $H_{2}(x) = -x\log_{2} (x)
- (1-x) \log_{2} (1- x)$, \ and $\xi(F) = \frac{1}{d} \left[
\sqrt{F} + \sqrt{(d-1)(1-F)}\right]^{2}$ \cite{TV00,FJ}. We have
chosen to use the entanglement of formation from among the various
entanglement measures
\cite{Wootters97,HHH96,RBC01,TV00,FJ,ZHHH01,HSR03,CMB04,DB04}. Less
standard measures, such as concurrence and negativity, have
typically been used to study ESD. The concurrence is a readily
calculated mixed-state entanglement measure for $2\times2$ systems
\cite{Wootters97}. The negativity can be used for mixed states of
$2\times 2$ and $2\times3$ systems \cite{HHH96,AJ072}. For larger
finite-dimensional bipartite systems, there is no known general
closed form expression for entanglement applicable to all states.
However, we can use the above specific form for the entanglement of
formation that is valid for arbitrary isotropic mixed states of such
systems, our case. Eq. 2 is valid for $d>2$ (although it does not
apply in the case $d=2$); Terhal and Vollbrecht showed its validity
for $d=3$ and conjectured it for arbitrary $d\geq 3$ \cite{TV00}, a
conjecture later proven to be true by Fei and Li-Jost \cite{FJ}.

For ESD to occur, entanglement must be initially positive and go to
zero in finite time. To demonstrate that ESD from an isotropic
initial state $\rho_{\rm iso}(d)$, it suffices to show that the
fidelity $F\left(\rho_{\rm iso}(d),P(\ket{\Psi(d)}\right)$ is
initially above $F_{\rm critical}= d^{-1}$ and later drops to that
value at some $t<\infty$. Our interest is in states of qudit pairs
with $d>2$. We begin with a simple model illustrating basic
dephasing, based on which conclusions about the general case of
isotropic noise, wherein initially isotropic states are certain to
remain isotropic, are later drawn. The general time-evolved
open-system density matrix expressible in the operator-sum
decomposition of an open-system evolution is the completely positive
trace preserving (CPTP) map $\rho\left(t\right) =
\mathcal{E}\left(\rho\left(0\right)\right) =
\sum_{\mu}K_{\mu}^{\dagger}\left(t\right)\rho\left(0\right)
K_{\mu}\left(t\right)$; the operators $\{K_\mu(t)\}$ satisfy the
completeness condition $\sum_{\mu}K_{\mu}^{\dagger}(t)K_{\mu}(t) =
\mathbb{I}$ and the trace preserving condition
$\sum_{\mu}K_{\mu}(t)K_{\mu}^{\dagger}(t) = \mathbb{I}$, and
represent the influence of statistical noise \cite{NC,YE02,YE03}.
For our model of multi-local dephasing noise acting on a bipartite
state,  $K_{\mu}(t) = D_{j}(t)E_{i}(t)$:
\begin{eqnarray}
\rho\left(t\right) = \mathcal{E}\left(\rho\left(0\right)\right) =
\sum_{i,j = 1}^{2}
D_{j}^{\dagger}\left(t\right)E_{i}^{\dagger}\left(t\right)
\rho\left(0\right)
E_{i}\left(t\right)D_{j}\left(t\right)\label{krausGeneral} ,
\end{eqnarray}
where $E_{i}(t)$ and $D_{j}(t)$ correspond to local dephasing noise
components acting on the first and second qudit, respectively, and
individually satisfy the above conditions. We take these to be of
the specific forms
\begin{eqnarray}
E_{1}(t) &=&
{\rm diag}(1,\gamma_{\rm A},\gamma_{\rm A},\ldots,\gamma_{\rm A}) \otimes \mathbb{I}_{d} \ ,
E_{2}(t)  =
{\rm diag}(0,\omega_{\rm A},\omega_{\rm A},\ldots,\omega_{\rm A}) \otimes \mathbb{I}_{d} \ , \\
D_{1}(t) &=&
\mathbb{I}_{d} \otimes {\rm diag}(1,\gamma_{\rm B},\gamma_{\rm B},\ldots,\gamma_{\rm B}) \ ,
D_{2}(t)  = \mathbb{I}_{d} \otimes {\rm diag}(0,\omega_{\rm
B},\omega_{\rm B},\ldots,\omega_{\rm B})\ ,
\end{eqnarray}
where $\gamma_{\rm A}\left(t\right) = e^{-\Gamma_{{\rm A}}t/2}, \
\gamma_{\rm B}\left(t\right) = e^{-\Gamma_{{\rm B}}t/2}, \
\omega_{\rm A}\left(t\right) = \sqrt{1-\gamma_{{\rm A}}^{2}(t)},\
{\rm and} \ \omega_{\rm B}\left(t\right) = \sqrt{1-\gamma_{{\rm
B}}^{2}(t)}.$ For simplicity, these noise parameters are chosen so
that the rate of dephasing from state $k$ relative to the state 1
are equal, that is, $\Gamma_{\rm A} = \Gamma_{\rm B} = \Gamma$, and
hence $\gamma_{\rm A}(t) = \gamma_{\rm B}(t) = \gamma(t)$, although
subscripts may occasionally appear for clarity and the
time-dependence of $\gamma$'s may be implicit. This simple model
generalizes well to the case where dephasing occurs between all
states of our basis.

The initial value  $F_0$ of the time-dependent fidelity
$F\left(\rho(d,t), P(\ket{\Psi(d,t)}) \right) = {\rm tr}\left(
\rho(d,t) P(\ket{\Psi(d,t)}) \right)$ of the time-evolved states,
for each value of $d$, has an corresponding to the choice of initial
isotropic state, $\rho(d,0)$. The initial state is
\begin{equation}
\rho(d,0)= \epsilon\mathbb{I}_{d^{2}} + \zeta P(\ket{\Psi(d)})\ ,
\end{equation}
where $\epsilon\equiv\frac{1 - F_{0}}{d^{2} - 1}$ and $\zeta\equiv
\frac{F_{0} d^{2}-1}{d^{2}-1}$. The first term contributes a summand
of $\epsilon$ to each element of the density matrix diagonal and
nothing elsewhere, since it is a multiple of $\mathbb{I}_{d^{2}}$;
the second term, which involves $P(\ket{\Psi(d)})$, contributes
$\zeta d^{-1}$ at positions $\left({\rm row, col}\right)$ =
$\left((j-1)d+j,(k-1)d+k\right)$ for $1 \leq j, k \leq d$ and zeros
elsewhere. Here, the joint-system density matrix is studied in the
tensor product of the individual subsystem bases $\ket{1} = \left(1
0 \ldots 0\right)^{\rm T}, \ket{2} = \left(0 1 0 \ldots
0\right)^{\rm T}, \ldots, \ket{d} = \left(0\ldots 0 1\right)^{\rm
T}$. The initial state density matrix in explicit matrix form is
\begin{eqnarray}
\rho(d,0) &=& \epsilon\mathbb{I}_{d^{2}} + \zeta P(\ket{\Psi(d)}) \\
&=& {\rm diag}(\epsilon, \epsilon, \ldots, \epsilon) +
\frac{1}{d}\left(
\begin{array}{cccc}
\mathcal{M}_{I}   & \cdots & \mathcal{M}_{I}   & \mathcal{M}_{II}  \\
\mathcal{M}_{I}   & \cdots & \mathcal{M}_{I}   & \mathcal{M}_{II}  \\
\vdots            & \vdots & \vdots            & \vdots            \\
\mathcal{M}_{I}   & \cdots & \mathcal{M}_{I}   & \mathcal{M}_{II}  \\
\mathcal{M}_{III} & \cdots & \mathcal{M}_{III} & \mathcal{M}_{IV} \\
\end{array}
\right)\ ,
\end{eqnarray}

\begin{equation}
\mathcal{M}_{I} = \left(
\begin{array}{ccccc}
 \ \zeta \ \   & \ 0 \ \   & \cdots & \ 0 \ \        & \ 0 \ \       \\
0      & 0      & \cdots      & 0      & 0      \\
\vdots & \vdots & \vdots & \vdots & \vdots \\
0      & 0      & \cdots      & 0      & 0      \\
0      & 0      & \cdots & 0      & 0
\end{array}
\right)\ , \ \mathcal{M}_{II} = \left(
\begin{array}{ccccc}
\ \zeta \ \  & \ 0 \ \      & \cdots & \ 0 \ \      & \ \zeta \ \  \\
0      & 0      & \cdots      & 0      & 0      \\
\vdots & \vdots & \vdots & \vdots & \vdots \\
0      & 0      & \cdots      & 0      & 0      \\
0      & 0      & \cdots & 0      & 0      \\
\end{array}
\right)\ , \nonumber \\
\end{equation}
\begin{equation}
\mathcal{M}_{III} = \left(
\begin{array}{ccccc}
 \ \zeta \ \  & \ 0 \ \      & \cdots & \ 0 \ \      & \ 0 \ \     \\
 0      & 0      & \cdots      & 0      & 0      \\
 \vdots & \vdots & \vdots & \vdots & \vdots \\
 0      & 0      & \cdots      & 0      & 0      \\
 \zeta  & 0      & \cdots & 0      & 0
\end{array}
\right)\ ,\ \mathcal{M}_{IV} = \left(
\begin{array}{ccccc}
 \ \zeta \ \  & \ 0 \ \      & \cdots & \ 0 \ \      & \ \zeta \ \  \\
 0      & 0      & \cdots      & 0      & 0      \\
 \vdots & \vdots & \vdots & \vdots & \vdots \\
 0      & 0      & \cdots      & 0      & 0      \\
 \zeta  & 0      & \cdots & 0      & \zeta
\end{array}
\right)\nonumber\ ,
\\
\end{equation}
wherein there are $d-2$ of the $\mathcal{M}_{I}$ $(d+1) \times
(d+1)$ Hermitian matrices on the rows and columns, $d-2$ of the
$\mathcal{M}_{II}$ $(d+1) \times (d+2)$ on the last column, $d-2$ of
the $\mathcal{M}_{III}$ $(d+2) \times (d+1)$ on the last row,
$\mathcal{M}_{IV}$ is a $(d+2) \times (d+2)$ Hermitian matrix;
$\mathcal{M}_{II} = \mathcal{M}_{III}^{\dagger}$.

The time-evolved density matrix $\rho\left(d,t\right) =
\mathcal{E}\left(\rho\left(d,0\right)\right)$, that is, the solution
of Eq. \ref{krausGeneral} for $t\geq 0$, consists of decaying
factors $\tilde{\gamma}(t)$ multiplying the elements of $\rho(d,0)$
at $\left({\rm row, 1}\right)$ = $\left((j-1)d+j,1\right)$ for $2
\leq j \leq d$ and at $\left({\rm 1, col}\right)$ =
$\left(1,(k-1)d+k\right)$ for $2 \leq k \leq d$, where
$\tilde{\gamma}(t)$ represents $\gamma_{\rm A}(t)$, $\gamma_{\rm
B}(t)$, and $\gamma_{\rm A}(t)\gamma_{\rm B}(t)$ in the cases of
local noise acting on A alone, B alone, and on both, respectively;
that is, decaying terms appear in the first row and first column
only, because in the simple noise model we consider for now there is
dephasing of the $k^{th}$ state for $2 \leq k \leq d$ relative to
the ground state $k=1$, but no dephasing between other basis states.
Because we are not concerned with precisely when full
disentanglement occurs, only that it \emph{does} occur in finite
time, specific decay rates appearing in the $\tilde{\gamma}(t)$ and
from hereon collectively designated $\tilde{\Gamma}$, are not
crucial---they must only be nonzero. The time-dependent state is
\begin{eqnarray}
\rho(d,t) &=& \epsilon\mathbb{I}_{d^{2}} + \zeta P(\ket{\Psi(d,t)}) \\
&=& {\rm diag}(\epsilon, \epsilon, \ldots, \epsilon) +
\frac{1}{d}\left(
\begin{array}{ccccc}
\mathcal{M}_{I}                & \mathcal{M}_{I}\tilde{\gamma}& \cdots & \mathcal{M}_{I}\tilde{\gamma}   & \mathcal{M}_{II}\tilde{\gamma}  \\
\mathcal{M}_{I}\tilde{\gamma}  & \mathcal{M}_{I} & \cdots & \mathcal{M}_{I}   & \mathcal{M}_{II}  \\
\vdots                         & \vdots & \vdots & \vdots                          & \vdots            \\
\mathcal{M}_{I}\tilde{\gamma}  & \mathcal{M}_{I} & \cdots & \mathcal{M}_{I}   & \mathcal{M}_{II}  \\
\mathcal{M}_{III}\tilde{\gamma}& \mathcal{M}_{III}& \cdots & \mathcal{M}_{III} & \mathcal{M}_{IV} \\
\end{array}
\right)\ .
\end{eqnarray}
The bipartite system state will remain partially coherent for all
finite times because all off-diagonal elements persist for all
finite times; only in the limit $t\rightarrow\infty$ is there full
decoherence between the ground state and every other state. However,
as we now show, there still is complete loss of entanglement in
\emph{finite} time for a range of initial isotropic states. It is
valuable to note here that the production of such states and their
non-local measurement may be experimentally challenging.

To see that complete disentanglement does indeed take place in
finite time, we first find the time-dependent fidelity
$F\left(\rho(d,t),P(\ket{\Psi})\right) = {\rm
tr}\left(\rho(d,t),P(\ket{\Psi})\right)$. The argument
$\rho(d,t)P(\ket{\Psi})=\bm{M}$ has three distinct sorts of terms,
C$_{1}$, C$_{2}$, and C$_{3}$, having specific forms which we
describe in turn and then evaluate. The sole C$_{1}$ term appears at
$\bm{M}_{1,1}$; C$_{2}$ terms appear at $\bm{M}_{\rm row, col}$ with
$\left({\rm row, col}\right)$ = $\left((j-1)d + j, (k-1)d +
k\right)$ for $2 \leq j, k \leq d$, $\delta_{jk} = 1$; C$_{3}$
consists of the remaining terms of the matrix. We designate the
values of the terms of sorts $C_{1}$, $C_{2}$, and $C_{3}$, by
$c_1$, $c_2,$ and $c_3$, respectively. The fidelity has nontrivial
contributions only from terms from the first and second of these
classes, of which there are numbers $N_1$ and $N_2$, respectively.
In the above simple model, C$_{1}$ consists of the single term
appearing as $\bm{M}_{1,1}$, being the inner product of the first
row of $\rho(d,t)$ and the first column of
$P\left(\ket{\Psi(d,t)}\right)$, taking the value $c_{1} =
\left(\epsilon+\frac{\zeta}{d}\right)\left(\frac{1}{d}\right) +
\left(
\frac{\zeta}{d}\tilde{\gamma}(t)\right)\left(\frac{1}{d}\right)\left(d-1\right)$,
and $N_1=1$. C$_{2}$ terms are those appearing at $\bm{M}_{\rm row,
col}$ for $\left({\rm row, col}\right)$ = $\left((j-1)d+j, (k-1)d+k
\right)$ for $2 \leq j, k \leq d$ with $\delta_{jk} = 1$, and are
inner products, each taking the value $c_{2} =
\left(\frac{\zeta}{d}\tilde{\gamma}(t)\right)\left(\frac{1}{d}\right)
+ \left(\epsilon+\frac{\zeta}{d}\right)\left(\frac{1}{d}\right) +
\left(\frac{\zeta}{d}\right)\left(\frac{1}{d}\right)\left(d-2\right)$,
and  $N_2=d-1$. The time-dependent fidelity for this model is thus
\begin{equation}
F\left(\rho(d,t),P(\ket{\Psi(d,t)}) \right) = c_{1}N_{1} +
c_{2}N_{2} + c_{3}N_{3} =2\frac{(d^{2}F_0-1)\tilde{\gamma}(t) +
d^2(d-1)\frac{F_0}{2} + 1}{d^3+d^2}\ ,
\end{equation}
which is determined by the initial state fidelity $F_{0}$, $d$ of
the individual qudits, and $t$.

Recall that $F(d,t)$ calculated above must initially be above the
value $F_{\rm critical}(d) = d^{-1}$ at and below which isotropic
states are separable, that is, the entanglement of formation is
zero, and \emph{in finite time} reach that value in order for
entanglement sudden death to occur. Note that separability occurs
whenever the entanglement is zero independently of the particular
entanglement measured used, because this is a defining property any
valid entanglement measure. Considering now $G(d,t)\equiv F(d,t) -
F_{\rm critical}(d)$, we show that both $F_0(d)>F_{\rm critical}(d)$
and this function $G(d,t)=0$ for some $t < \infty$, for a specific
form of $F_0(d)$. Taking the initial fidelities to be $F_0(d)
=(d-1)^{-1}$, we have in this simple model
\begin{equation}
G(d,t)\big|_{F_0(d)} = F(d,t)\big|_{F_0(d) =(d-1)^{-1}} - F_{\rm
critical}(d) =\frac{2(d^{2}-d+1)\tilde{\gamma}(t)-(d-1)(d-2)}
{d^{2}(d^{2} - 1)}\ ,
\end{equation}
which is $[d(d-1)]^{-1}>0$ at $t=0$ and is zero at time
$t=(2/\tilde{\Gamma})\ln[2(d^2-d+1)/(d-1)(d-2)]$. Recall that in the
noise model considered thus far, dephasing noise occurs only between
the ground state $k=1$ and the $k^{th}$ basis state (for $k = 2, 3,
\ldots, d$). This model is neither the simplest case of local
dephasing, wherein there is dephasing between only two particular
local basis states, nor is it the most general case wherein
dephasing occurs between all pairs of states within each subsystem.
Under it, initially isotropic states become anisotropic. However,
the expressions resulting from this noise simply generalize to the
case of the noise model inducing dephasing between \emph{all} local
basis states, in which isotropic states remain isotropic, that is
descriptive of what would be encountered in a highly random local
phase-noise environment: the solution for the time-dependent density
matrix differs from the above solution only by a $\tilde{\gamma}(t)$
decay factor in \emph{each} nonzero off-diagonal element. Because
the dephasing noise is isotropic in this general case, the
time-evolved states remain isotropic and the resulting fidelity
$F(\rho(d,t),P(|\Psi(d,t)\rangle))$ properly determines the
entanglement.

The terms of the C$_{1}$ and C$_{3}$ types contributing to the
fidelity are unchanged under this generalization, but the
C$_{2}$-type terms change:
$c_{2}=
\left(\frac{\zeta}{d}\tilde{\gamma}(t)\right)\left(\frac{1}{d}\right)
+ \left(\epsilon+\frac{\zeta}{d}\right)\left(\frac{1}{d}\right) +
\left(\frac{\zeta}{d}\right)\left(\frac{1}{d}\right)\left(d-2\right)
 \rightarrow
\left(\frac{\zeta}{d}\tilde{\gamma}(t)\right)\left(\frac{1}{d}\right)
+ \left(\epsilon+\frac{\zeta}{d}\right)\left(\frac{1}{d}\right) +
\left(\frac{\zeta}{d}\tilde{\gamma}(t)\right)\left(\frac{1}{d}\right)\left(d-2\right)$.
An ``additional'' factor of $\tilde{\gamma}(t)$ appears in the third
contribution. The effect on the functions $F(d,t)$ and $G(d,t)$ of
this extra decay factor is only a more rapid decrease because the
third term also decays to zero. There is no qualitative effect on
the behavior of $F$ and $G$: the fidelity only decreases more
rapidly. However, the resulting fidelity now determines the
entanglement $E_{\rm f}$. Hence, ESD occurs for qudit-qudit systems
for all finite qudit-space dimensions $d$ greater than $2$, when
initially prepared in appropriate entangled isotropic states subject
to dephasing noise alone. It continues to be exhibited for values of
$d$ large like the infinite-dimensional bipartite systems studied in
\cite{DH04}.


\begin{references}
\bibitem{NC} Nielsen, M. A., and I. L. Chuang, \emph{Quantum
computation and quantum information}\\ (Cambridge University Press;
Cambridge, 2000).
\bibitem{Preskill} J. Preskill, Proc. Roy. Soc. London A {\bf 454}, 385 (1998).
\bibitem{YE02} T. Yu and J. H. Eberly, Phys. Rev. B {\bf 66}, 193306 (2002).
\bibitem{YE03} T. Yu and J. H. Eberly, Phys. Rev. B {\bf 68}, 165322 (2003).
\bibitem{YE04} T. Yu and J. H. Eberly, Phys. Rev. Lett. {\bf 93}, 140404 (2004).
\bibitem{AJ07} K. Ann and G. S. Jaeger, Phys. Rev. B, {\bf 75}, 115307 (2006).
\bibitem{AJ072} K. Ann and G. S. Jaeger, Phys. Lett. A (in press;
doi:10.1016/j.physleta.2007.07.070).
\bibitem{DH04} P. J. Dodd and J. J. Halliwell, Phys. Rev. A {\bf 69}, 052105 (2004).
\bibitem{YE06} T. Yu and J. H. Eberly, Phys. Rev. Lett. {\bf 97}, 140403 (2006).
\bibitem{CLR04}  C. F. Roos, G. P. T. Lancaster, M. Riebe, H. Haffner, W. Hansel, S. Gulde, C. Becher, J. Eschner, F. Schmidt-Kaler, R. Blatt, Phys. Rev. Lett. {\bf 92}, 220402 (2004).
\bibitem{ADH07}  M. P. Almeida, F. de Melo, M. Hor-Meyll, A. Salles, S. P. Walborn, P. H. Souto Ribeiro, L.
Davidovich, Science {\bf 316}, 579 (2007).
\bibitem{LCD07}  J. Laurat, K. S. Choi, H. Deng, C. W. Chou, H. J. Kimble, arxiv:0706.0528.
\bibitem{Jozsa} R. Jozsa, J. Mod. Opt. {\bf 41}, 2315 (1994).
\bibitem{HH99} M. Horodecki and P. Horodecki, Phys. Rev. A {\bf 59}, 4206 (1999).
\bibitem{FJ} S.-M. Fei and X. Li-Jost, Phys. Rev. A {\bf 73}, 024302 (2006).
\bibitem{TV00}B. M. Terhal and KarlGerd H. Vollbrecht, Phys. Rev. Lett. {\bf 85}, 2625 (2000).
\bibitem{ZHHH01} K. Zyczkowski, P. Horodecki, M. Horodecki, R. Horodecki, Phys. Rev. A {\bf 65}, 012101 (2001).
\bibitem{HSR03}  M. Horodecki, P. W. Shor, M. B. Ruskai, Rev. Math. Phys. {\bf 15}, 629 (2003).
\bibitem{CMB04}  A. R. R. Carvalho, F. Mintert, A. Buchleitner, Phys. Rev. Lett. {\bf 93}, 230501 (2004).
\bibitem{DB04}   W. Dur, H.-J. Briegel, Phys. Rev. Lett. {\bf 92}, 180403 (2004).
\bibitem{Wootters97} W. K. Wootters, Phys. Rev. Lett. {\bf 80}, 2245 (1997).
\bibitem{HHH96} M. Horodecki, P. Horodecki, R. Horodecki, Phys. Lett. A 223, 1 (1996).
\bibitem{RBC01} P. Rungta, V. Buzek, C. M. Caves, M. Hillery, G. J. Milburn, Phys. Rev. A {\bf 64}, 042315 (2001).
\end{references}
\end{document}